\pdfoutput=1

\documentclass[a4paper]{jpconf}
\usepackage{graphicx}

\begin{document}
\title{Stable operation with gain of a double phase Liquid Argon LEM-TPC with a 1~mm thick segmented LEM}

\author{A.~Badertscher, A.~Curioni, S.~Horikawa, L.~Knecht, D.~Lussi, A.~Marchionni, G.~Natterer, F.~Resnati, A.~Rubbia, T.~Viant}

\address{Institute for Particle Physics, ETH Zurich, 8093 Zurich, Switzerland}

\ead{andre.rubbia@cern.ch}

\begin{abstract}
In this paper we present results from a test of a small Liquid Argon Large Electron Multiplier Time Projection Chamber (LAr LEM-TPC).
This detector concept provides a 3D-tracking and calorimetric device capable of charge 
amplification, suited for next-generation neutrino detectors and possibly direct Dark Matter searches.
During a test of a 3~lt chamber equipped with a 10$\times$10~cm$^2$ readout,
cosmic muon data was recorded during three weeks of data taking.
A maximum gain of 6.5 was achieved and the liquid argon was kept pure 
enough to ensure 20~cm drift (O(ppb)~O$_2$ equivalent).
\end{abstract}

\section{Introduction}
The Liquid Argon Large Electron Multiplier Time Projection Chamber (LAr LEM-TPC) is a novel kind of double phase (liquid-vapor) noble gas TPC 
suited for the realization of next generation detectors for neutrino physics and proton decay searches~\cite{Rubbia04}, considered as well as for direct 
Dark Matter detection~\cite{Rubbia06}. The ionization charge is drifted towards the liquid-vapor interface, where it is extracted by means of an appropriate
electric field produced by two grids. In the vapor phase, Townsend avalanche takes place in the high electric field regions confined 
in the LEM holes, similar to the situation of the Gas Electron Multiplier (GEM)~\cite{Sauli97}. The charge is collected by a set
of segmented electrodes, on which the induced signals are recorded.
The resulting charge multiplication yields a significant improvement of the $S/N$ ratio compared to the single phase liquid argon TPC~\cite{Badertscher09}. 
The very high quality of the images and the $dE/dx$ measurement are key points to address the three-dimensional geometry of the ionizing events and to identify the involved particles. In addition the charge amplification technique is compatible with very long drifts, because it allows to compensate for
the diffusion of the electron cloud and the potential charge loss due to electronegative impurities diluted in the liquid argon.

The detector described in this paper was developed during the last three years and different tests of the charge readout devices were performed~\cite{Badertscher09,Badertscher08}. Here we present the results obtained with a 1~mm thick Large Electron Multiplier (LEM) with an area of 10$\times$10~cm$^2$.

\section{Description of the detector}
\begin{figure}[h]
\begin{minipage}[b]{18pc}
\begin{center}
\includegraphics[height=20pc]{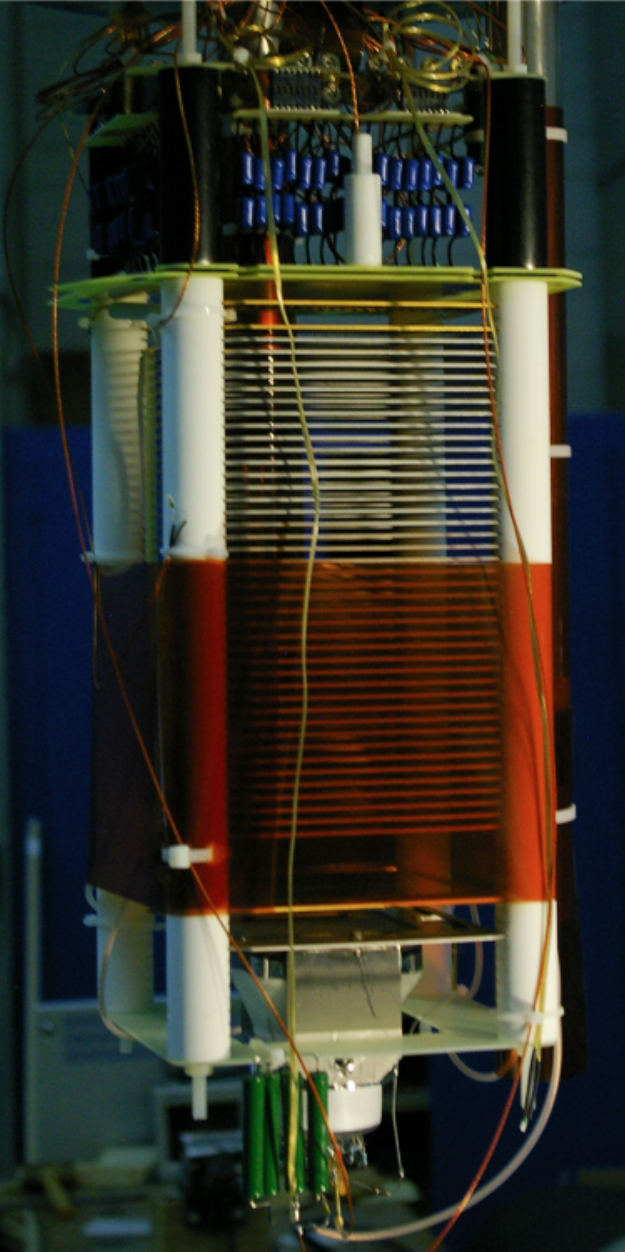}
\end{center}
\caption{\label{picture}Picture of the detector.}
\end{minipage}
\hfill
\begin{minipage}[b]{18pc}
\begin{center}
\includegraphics[height=16pc]{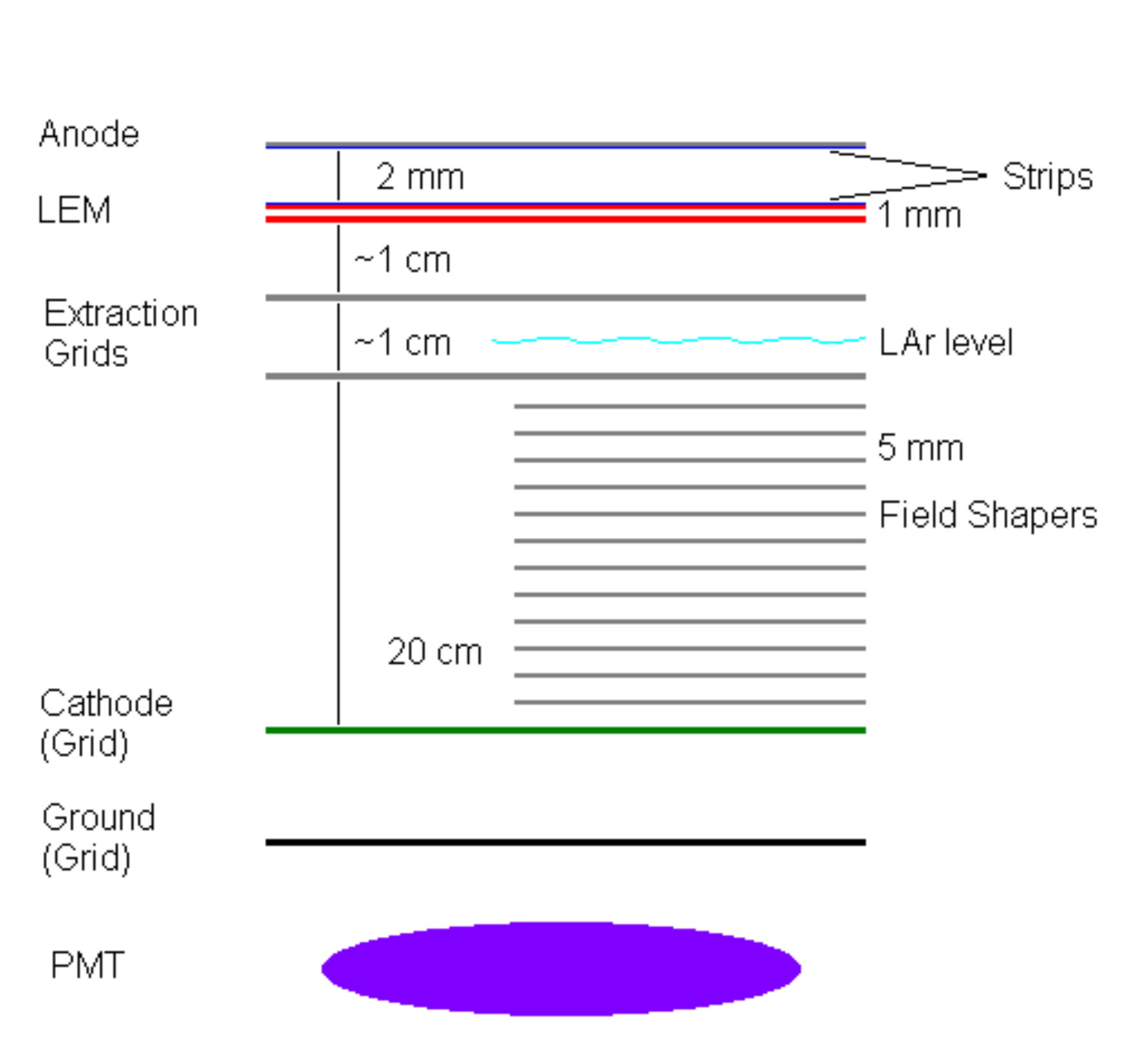}
\end{center}
\caption{\label{scheme}Schematic representation of the detector.}
\end{minipage}
\end{figure}

The picture of the detector is presented in figure~\ref{picture}. A set of field shapers, spaced every 5~mm, defines the drift volume, 20~cm long and with an 
area of 10$\times$10~cm$^2$.
An electric field of the order of 0.5-1.0~kV/cm is applied to drift electrons vertically towards the liquid surface.
The electrons are extracted from the liquid to the vapor phase by means of a set of two grids parallel to the liquid surface. Each grid is made out of 100~$\mu$m diameter stainless steel wires, spaced 5~mm. During operation the level of the liquid argon is precisely set in between the two grids and the voltages are tuned to obtain an electric field of about 3~kV/cm just below the liquid surface (see the schematics in figure~\ref{scheme}).

The charge readout apparatus (LEM and anode) is installed in the vapor phase 1~cm above the top grid.
The LEM is an electron multiplier with millimeter-size holes and a thickness of about 1~mm. 
It is a coarser and more robust extrapolation of the Gas Electron Multiplier (GEM)~\cite{Sauli97}. 
The LEM is produced with standard Printed Circuit Board (PCB) techniques and the holes are mechanically drilled through the PCB plate. The drifting electrons are driven into the LEM holes where a high electric field (larger than 25~kV/cm) allows to trigger the Townsend avalanche -- a detailed description of the working principle can be found in~\cite{Badertscher08}. When the amplified charge leaves the holes, it is collected partially on the anode, installed 2~mm above the LEM, and partially on the top electrode of the LEM. The sharing of the charge between the two electrodes can be adjusted changing the electric field between the anode and the LEM. Each electrode is segmented into 16~strips 6~mm wide and the two sets of strips are orthogonal to  allow the reconstruction of the X-Y coordinates of the ionizing event, while the Z coordinate is given by the drift velocity multiplied by the drift time. In table~\ref{geometryParameters} the geometry parameters of the LEM are summarized.

\begin{table}[h]
\begin{minipage}[b]{16pc}
\begin{center}
\begin{tabular}{lr}
  \hline
  \hline
  Material & copper-cladded FR4\\
  Active area & 10x10~cm$^2$ \\
  PCB thickness & 1.0~mm \\
  Copper thickness & 18~$\mu$m \\
  Hole diameter & 500~$\mu$m \\
  Hole rim & 100~$\mu$m \\
  Hole pitch & 800~$\mu$m \\
  Strip width & 6~mm \\
  \# strips & 16 \\
  \hline
  \hline
\end{tabular}
\end{center}
\end{minipage}
\hfill
\begin{minipage}[b]{16pc}
\caption{\label{geometryParameters}Geometry parameter of the described LEM.}
\end{minipage}
\end{table}

The detector is also equipped with a photomultiplier tube (PMT) that provides the time reference of the events and additional information about the purity of the argon, since the characteristic time of the radiative de-excitation of argon excimer states 
depends on the concentration (down to the ppm level) of impurities such as H$_2$O, O$_2$, N$_2$~\cite{Acciarri08:O2,Acciarri08:N2}.
A more sensitive way to monitor the purity of liquid argon is to measure the charge loss along the drift path (i.e. as a function of the drift time). To ensure a drift as long as 20~cm the liquid argon purity must be of the order of 1~ppb (oxygen equivalent) or less, since the oxygen content~([O$_2$]) is known to be related to the drifting electron lifetime~($\tau_{e^-}$) by $[\mbox{O}_2]\times\tau_{e^-}\approx300\mbox{ $\mu$s ppb}$~\cite{Buckley89}. Two methods are used to achieve the needed purity: (1)~during the filling procedure the liquid argon is filtered with a custom made cartridge loaded with reduced copper-oxyde powder and (2)~to maintain the purity during operation, liquid argon is evaporated, pushed by a bellows pump through a commercial SAES getter\footnote{SAES Pure Gas Inc., MicroTorr MC400.}, recondensed through a long serpentine immersed in a liquid argon open bath, and again reintroduced as liquid inside the detector. A proof of the effectiveness of the purification system is given in this paper. For a detailed description of the recirculation circuit see Ref.~\cite{Badertscher09}.

The detector main vessel was evacuated with a residual pressure smaller than 5$\times$10$^{-6}$~mbar in order to favour the outgassing of the materials and check the vacuum tightness of the system.
The filling procedure consists in recirculating pure argon gas through the getter while cooling down the vessel in an open bath of liquid argon. The presence of gas ensures a fast and uniform thermalization of the detector and the recirculation system traps the outgassed molecules into the getter.
Finally the detector is filled with liquid argon adjusting the level in the middle of the two extraction grids, that are used also as a capacitive level meter.

\section{Cosmic muon reconstruction and argon purity}
\begin{figure}[h]
\begin{minipage}[b]{24pc}
\begin{center}
\includegraphics[width=24pc, height=10pc]{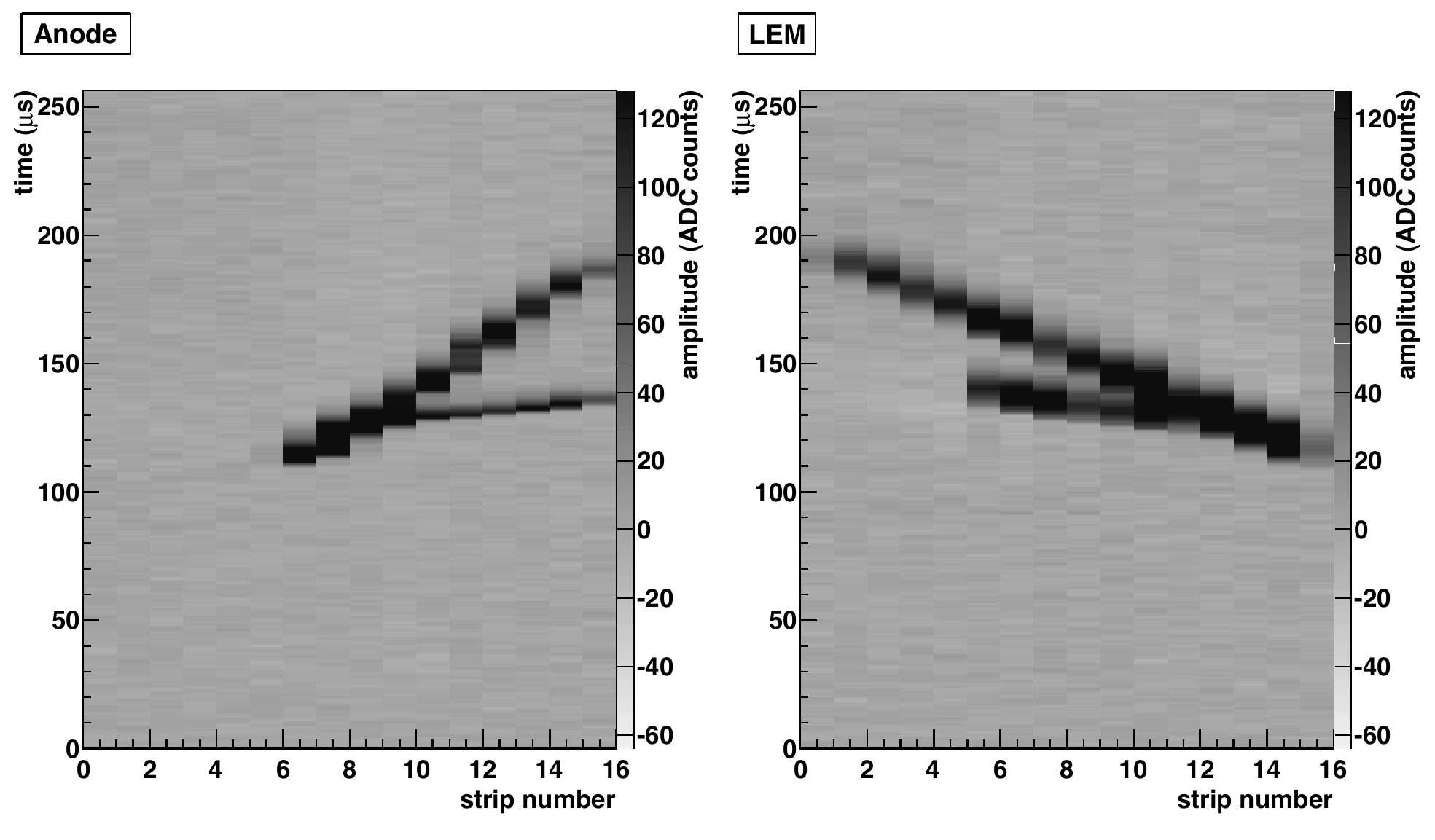}
\includegraphics[width=24pc, height=10pc]{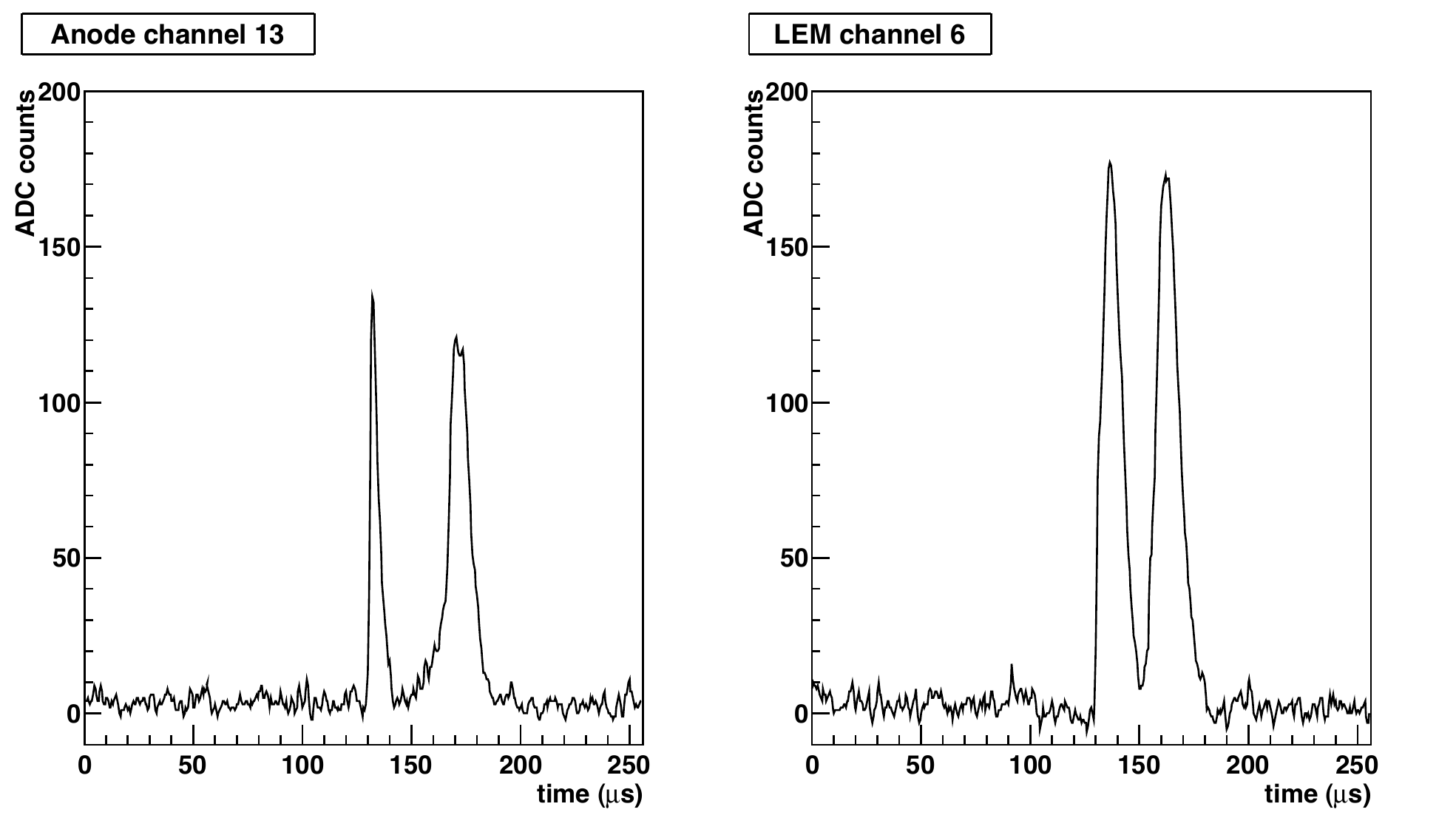}
\end{center}
\end{minipage}
\hfill
\begin{minipage}[b]{12pc}
\caption{\label{eventDisplay}Event display of a cosmic muon producing a delta ray. The two figures on top show the signals from the same event recorded on anode electrodes~(left) and on the LEM electrodes~(right). Below two examples of signal shapes (on the LEM strip number \#6 and anode \#13) are shown.}
\end{minipage}
\end{figure}

Cosmic muon events are used to characterize the detector because (1)~they release a defined amount of charge per unit path so they can be used as calibration source and (2) they cross the entire volume allowing the evaluation of the liquid argon purity.
A typical cosmic muon track with a delta ray emission is presented in figure~\ref{eventDisplay}, where the channel number and the drift time are plotted on the X-axis and Y-axis respectively and the gray scale is proportional to the signal amplitude, sampled every 400~ns (2.5~MHz). The two images refer to two orthogonal views of the same ionizing event. We also report as examples two signal waveforms, on the LEM strip number \#6 and anode \#13.

Cosmic muon tracks are used to determine the effective gain and the drifting electron lifetime.
Each event is three-dimensionally reconstructed in order to evaluate the track length~($\Delta$l) below each strip and the charge~($\Delta$Q) collected on each channel. The ratio
$\Delta$Q/$\Delta$l, proportional to the energy loss by the muon in liquid argon per unit length, is the relevant quantity used in the analysis.

\begin{figure}[h]
\begin{center}
\includegraphics[width=0.65\textwidth]{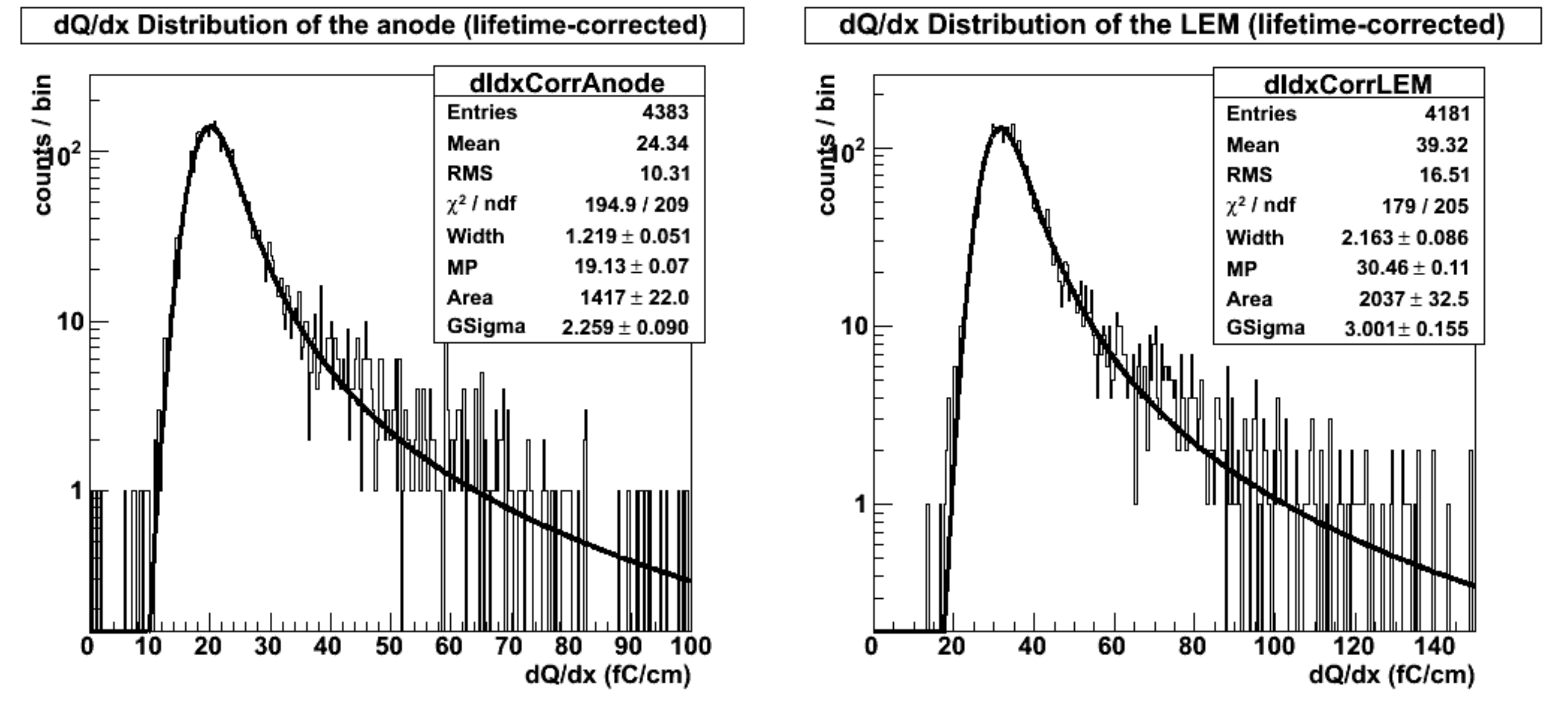}
\includegraphics[width=0.65\textwidth]{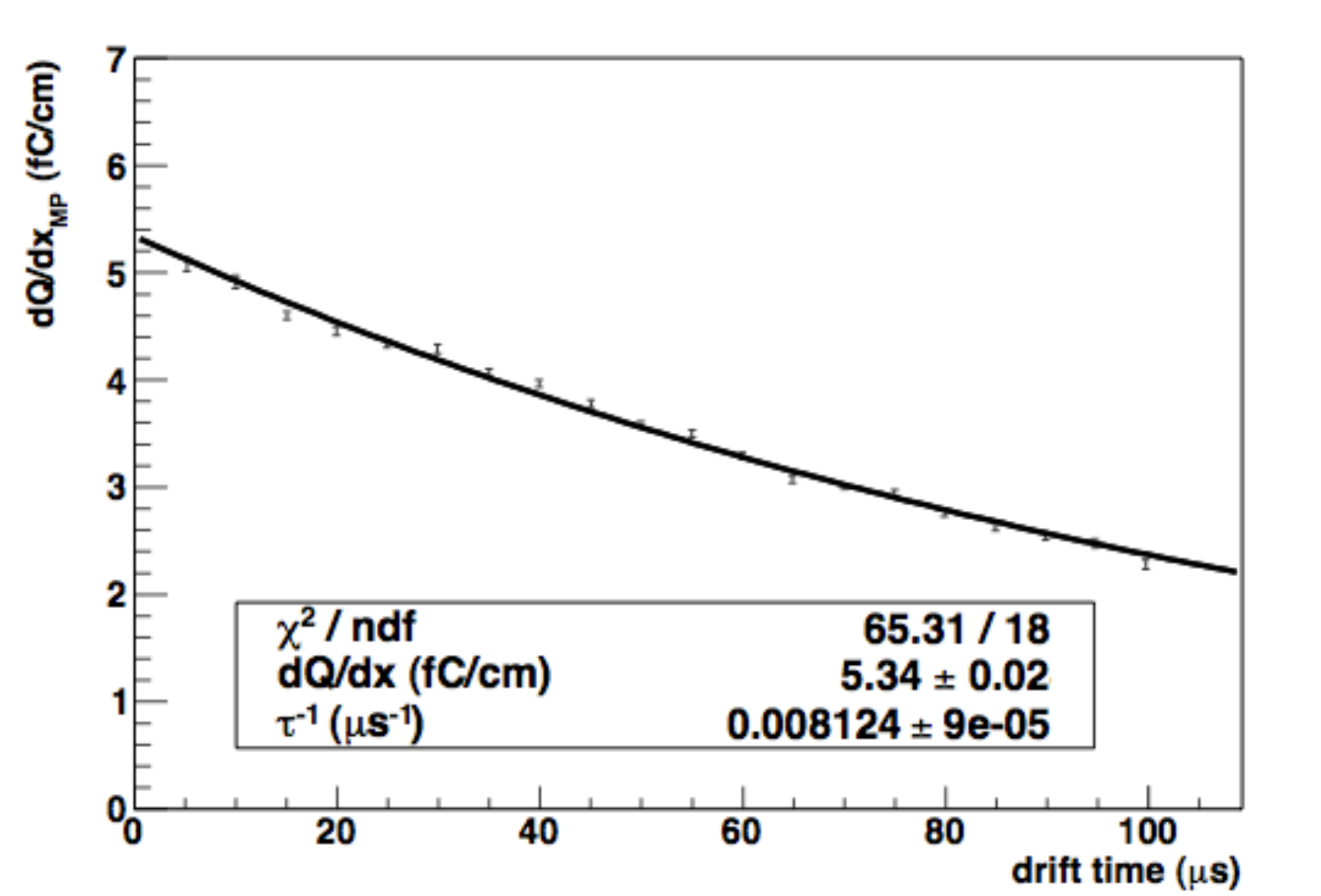}
\end{center}
\caption{\label{dQdxDistribution}Top: fit of the Gauss-convoluted Landau function to the $\Delta$Q/$\Delta$l distribution. Both the anode electrode view and the LEM electrode view are shown. Bottom: Most probable value evaluated from the fit of the Gauss-convoluted Landau function versus the drift time.}
\end{figure}

In order to evaluate the drifting electron lifetime, the drift volume is divided in equal time slices along the drift.
A Gauss-convoluted Landau function is fitted to the distribution of $\Delta$Q/$\Delta$l, as shown in figure~\ref{dQdxDistribution}.
The plot of the position of the Gauss-convoluted Landau maximum as a function of the drift time (drift slices) shows the exponential charge attenuation due to electronegative impurities.

\begin{figure}[htb]
\begin{center}
\includegraphics[width=0.65\textwidth]{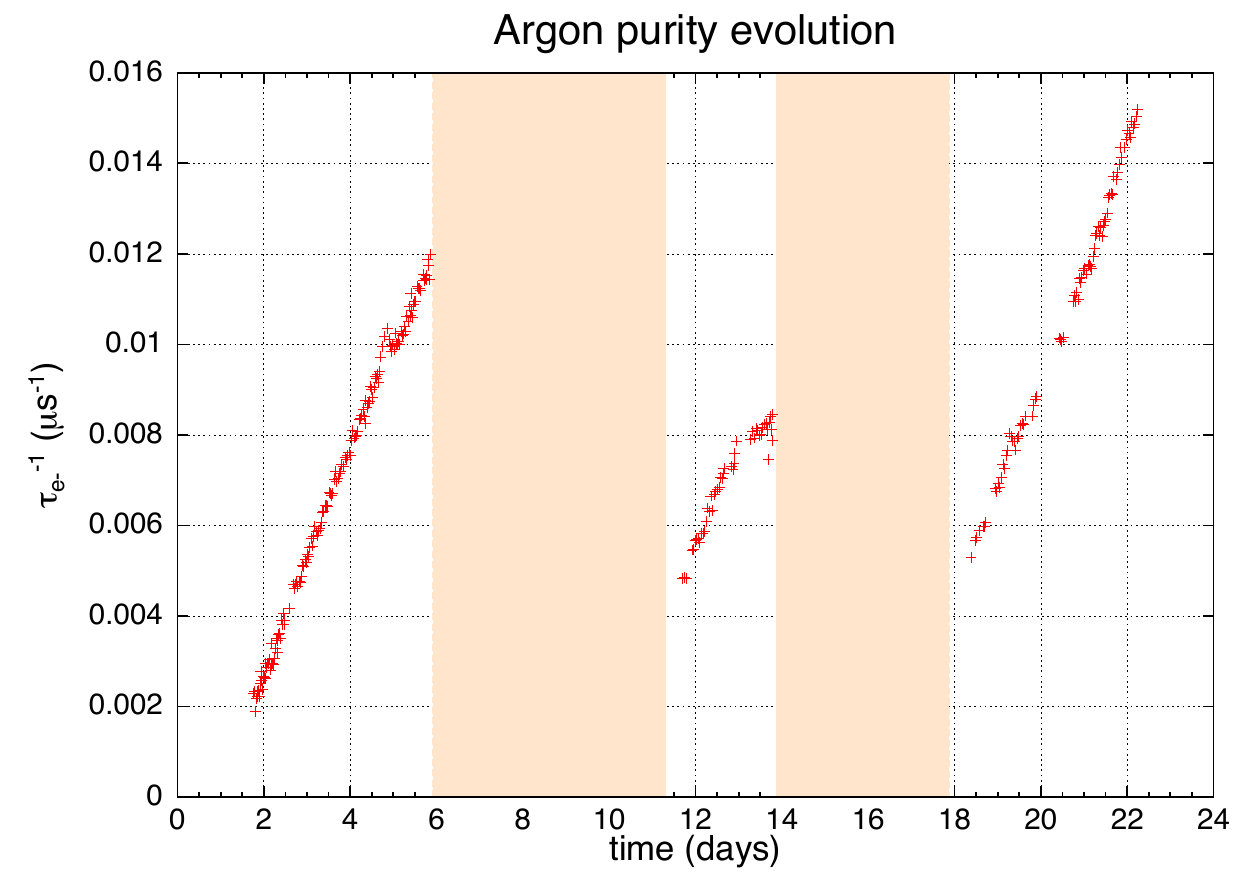}
\end{center}
\caption{\label{purityMonitor}History of the liquid argon purity. The value 1/$\tau_{e^-}$ is proportional to the impurities content. While acquiring data the purification system is off and the liquid argon purity degrades. During the purification (highlighted bands) the detector cannot be operated because the liquid argon wets the LEM and no amplification is possible.}
\end{figure}

The longest drifting electron lifetime (about 500~$\mu$s) is achieved at the beginning of the run. The corresponding O$_2$ equivalent content is about 0.6~ppb.
As shown in figure~\ref{purityMonitor} the liquid argon purity degrades with a rate of about 0.6~ppb/day, presumably due to the outgassing of the materials inside the detector or small leaks to the atmosphere. The purity is re-established by recirculating the argon gas through the purification cartridge and re-condensing the clean argon into the detector volume.

\section{Charge amplification and effective gain}
The {\it effective} gain is defined as the sum of the collected charge per unit length on the anode and LEM electrodes (corrected for the drifting electron lifetime) divided by the charge per unit length released in liquid argon by a minimum ionizing muon in an electric field of 500~V/cm. The former is estimated from the $\Delta$Q/$\Delta$l distribution of the reconstructed muon tracks and the latter is about 10~fC/cm, considering that a minimum ionizing muon in liquid argon releases 2.1~MeV/cm~\cite{Amerio04}.
The effective gain, in addition to the charge multiplication in the LEM holes, takes into account potential charge losses from the efficiency to extract the electrons to the vapor phase and from the transparency of the grids and the LEM (some electrons, in part due to the large diffusion in pure argon, may be collected on the grids or on the bottom electrode of the LEM). For this reason the effective gain is sensitive to the details of the electric field configuration. The working point is found from electrostatic computations of the drift lines and from the optimization of the charge sharing between the anode and the top LEM electrodes. The typical configuration of the electric field is reported in table~\ref{fieldConfiguration}.

\begin{table}[h]
\begin{minipage}[b]{16pc}
\begin{center}
\begin{tabular}{lr}
  \hline
  \hline
  Anode-LEM & 2~kV/cm\\
  LEM holes & 36~kV/cm\\
  Grid-LEM & 1~V/cm\\
  Extraction & 3~kV/cm\\
  Drift & 500~V/cm\\
  \hline
  \hline
\end{tabular}
\end{center}
\end{minipage}
\hfill
\begin{minipage}[b]{16pc}
\caption{\label{fieldConfiguration}Nominal configuration of electric fields for the run at 6.5 gain.}
\end{minipage}
\end{table}

\begin{figure}[h]
\begin{center}
\includegraphics[width=0.65\textwidth]{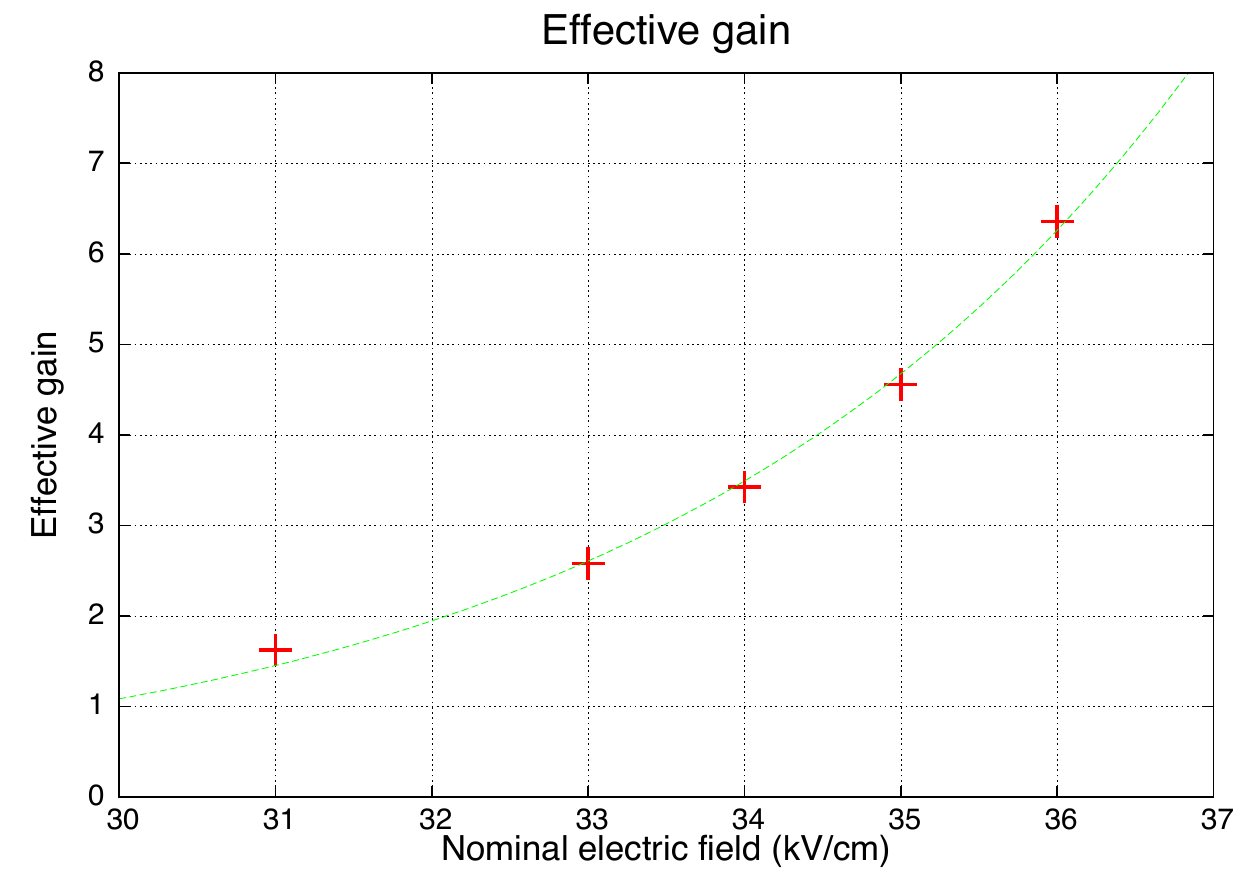}
\end{center}
\caption{\label{gainVsField}Effective gain as a function of the nominal electric field in the LEM holes. The curve is an exponential fit to the data to guide the reader's eyes.}
\end{figure}

Figure \ref{gainVsField} shows the effective gain as a function of the {\it nominal}\ electric field $E$ in the LEM holes, {\it defined} as the ratio $E=V/d$ of the potential difference $V$ applied to the LEM to the LEM thickness $d$\footnote{We note that the actual electric field is not constant along the path of electrons inside the holes, and the maximum electric field encountered by drifting electrons (plateau) is slightly less than the nominal electric field.}. The measurements were taken by varying the electric field $E$, while leaving all the other parameters unchanged, as reported in table~\ref{fieldConfiguration}.
Above 30~kV/cm we observe an effective gain larger than 1, but it is eventually limited at 6.5 by the occurrence of discharges, primarily 
in the LEM holes. The discharge pattern suggests that the voltage break-down are localized around regions of high electric fields, possibly close
to the strip segmentation of the upper electrode of the LEM. This observation led us to conclude that a promising alternative is to decouple 
the charge amplification stage from the readout stage with non-segmented LEM electrodes. Indeed,
such a configuration was successfully implemented after the completion of this work and improved gain
results will be published in a forthcoming paper~\cite{dlussiinprep}.
In the new setup, the X-Y readout feature was retained implementing a 2D~projective anode, based on the concept described
in Ref.~\cite{Bressan99}. 

\section{Conclusions}
In this paper we presented the successful and stable operation of a double phase LAr LEM Time Projection Chamber over a period of three weeks.
The gain was implemented with a 1~mm-thick LEM.
Cosmic muons,  recorded with very high quality imaging properties, were three-dimensionally reconstructed 
and used to evaluate the argon purity and the effective charge gain of the detector.
Thanks to a dedicated purification system, pure liquid argon could be obtained, reaching ppb  O$_2$ equivalent level  
in a few days, for drift paths of 20~cm.
The maximum effective gain achieved, measured directly from a fit to the energy deposited by the cosmic tracks, was about 6.5. 
A possibility to improve the gain performance is to avoid the segmentation of the top electrode of the LEM, since in this
configuration, break down discharges are often initiated in the high electric field regions separating the readout strips of the LEM.
The charge amplification greatly improves the imaging capabilities of the TPC making this charge readout technique interesting for 
next generation proton decay and neutrino experiments and possibly -- increasing the maximum gain -- for direct Dark Matter search experiments.

\end{document}